# Dynamic Cone-beam CT Reconstruction using Spatial and Temporal Implicit Neural Representation Learning (STINR)

## Running Title: Dynamic CBCT using STINR

You Zhang, Tielige Mengke

Advanced Imaging and Informatics in Radiation Therapy (AIRT) Laboratory
Medical Artificial Intelligence and Automation (MAIA) Laboratory
Department of Radiation Oncology
UT Southwestern Medical Center, Dallas, TX, 75235, USA

**Corresponding address:**

You Zhang, Ph.D., DABR

Department of Radiation Oncology

UT Southwestern Medical Center

Dallas, TX, 75235, USA

Phone: 919-627-3199

Email: you.zhang@utsouthwestern.edu

## Abstract

**Objective:** Dynamic cone-beam CT (CBCT) imaging is highly desired in image-guided radiation therapy to provide volumetric images with high spatial and temporal resolutions to enable applications including tumor motion tracking/prediction and intra-delivery dose calculation/accumulation. However, the dynamic CBCT reconstruction is a substantially challenging spatiotemporal inverse problem, due to the extremely limited projection sample available for each CBCT reconstruction (one projection for one CBCT volume).

**Approach:** We developed a simultaneous spatial and temporal implicit neural representation (STINR) method for dynamic CBCT reconstruction. STINR mapped the unknown image and the evolution of its motion into spatial and temporal multi-layer perceptrons (MLPs), and iteratively optimized the neuron weighting of the MLPs via acquired projections to represent the dynamic CBCT series. In addition to the MLPs, we also introduced prior knowledge, in form of principal component analysis (PCA)-based patient-specific motion models, to reduce the complexity of the temporal INRs to address the ill-conditioned dynamic CBCT reconstruction problem. We used the extended cardiac torso (XCAT) phantom to simulate different lung motion/anatomy scenarios to evaluate STINR. The scenarios contain motion variations including motion baseline shifts, motion amplitude/frequency variations, and motion non-periodicity. The scenarios also contain inter-scan anatomical variations including tumor shrinkage and tumor position change.

**Main results:** STINR shows consistently higher image reconstruction and motion tracking accuracy than a traditional PCA-based method and a polynomial-fitting based neural representation method. STINR tracks the lung tumor to an averaged center-of-mass error of <2 mm, with corresponding relative errors of reconstructed dynamic CBCTs <10%.

**Significance:** STINR offers a general framework that allows accurate dynamic CBCT reconstruction for image-guided radiation therapy. It is a one-shot learning method that does not rely on pre-training and is thus not susceptible to model generalizability issues. It also allows natural super-resolution. The framework can be readily applied to other imaging modalities as well.

# I. Introduction:

X-ray computed tomography (CT) is widely used in radiotherapy practices, providing volumetric images of high spatial resolution and geometric accuracy to guide radiotherapy planning and delivery [1-4]. Modern radiotherapy linear accelerators (LINACs) are commonly equipped with on-board X-ray imaging sources and flat panel detectors, which can acquire pre-delivery cone-beam CTs (CBCTs) for patient setup, plan adaptation, and dose accumulation [5-9]; and intra-delivery CBCTs for treatment positioning verification [10]. To acquire a fully-sampled CBCT image, the LINAC gantry needs to rotate at least $200°$ for a full-fan acquisition, or $360°$ when the detector is offset to increase the axial field-of-view (half-fan mode) [11]. Considering potential collision risks between the gantry and the patients, currently, the gantry rotation speed is mostly limited to $6°$ per second (s), which requires a substantial image acquisition time on the order of 1 minute. For cardiac and respiratory motion-impacted anatomical sites including chest and upper abdominal regions, the slow imaging speed results in CBCTs being affected by motion artifacts, which manifest as blurred anatomies and poorly defined structure boundaries [12]. Such artifacts introduce substantial uncertainties to the localization of moving tumors and surrounding organs at risk (OARs) for radiotherapy planning and treatment. The motion-blurred CBCTs fail to capture the motion trajectories of the anatomies, and may substantially underestimate and under-dose the radiotherapy target volume [13]. To address the motion challenges, respiratory-correlated CBCT, also named four-dimensional CBCT (4D-CBCT), has been developed [12, 14-17]. The 4D-CBCT technique assigns each acquired cone-beam projection to a respiratory phase based on tracked surrogate motion signals (surface motion, diaphragm motion, etc.), with each phase corresponding to a motion state along an assumed periodical motion cycle. It then reconstructs a semi-static CBCT volume at each phase bin, and the CBCT volumes from all bins are stacked to represent the motion kinematics during the nominal, averaged motion cycle. To address the under-sampling issues caused by the retrospective phase sorting, the 4D-CBCT phase number is usually limited to <= 10. In addition, the cone-beam projections are often intentionally over-sampled in number at a cost of imaging dose and scan time, to ensure an adequate amount of projections exist in each phase bin after phase sorting [16, 18, 19]. To reduce the imaging dose and scan time, different reconstruction algorithms, based on various *a priori* assumptions and motion models, were also developed to use limited projections within each phase bin to reconstruct high-quality, artifact-free 4D-CBCT images [14, 15, 20-25].

However, the 4D-CBCT imaging technique is essentially built upon the assumption that anatomical motion is periodical and regular, such that the projections acquired at different angles and time stamps can be sorted into the same phase bin. Although the motion of the chest and upper abdominal regions of real patients presents cardiac and pulmonary function-related periodicity, irregular and non-periodic motions, like those with amplitude/frequency variations or baseline shifts, are commonly observed as well [26-29]. Such irregularity may lead to substantial intra-phase motion variations and strong residual motion artifacts after sorting [30]. The nominal cycle resolved by 4D-CBCT fails to capture the irregularity and non-periodicity, which may provide crucial information on motion statistics and trends to guide patient immobilization, setup, and treatment monitoring [31, 32]. The ultimate solution to such a challenge is time-resolved CBCT imaging, or *dynamic* CBCT [33-36]. Dynamic CBCT, in contrast to the phase-resolved 4D-CBCT, reconstructs a continuous time series of volumetric images reflecting the spatial and temporal kinematics of patient anatomy without the phase-binning process. Dynamic CBCT essentially treats each CBCT projection as an individual phase and reconstructs a CBCT volume out of every single projection. However, the extreme under-sampling challenges the current reconstruction methods, as they require at least tens or hundreds of projections spanning over a large scan angle to reconstruct a high-quality volume. Some previous studies tried to address the ill-posed spatiotemporal dynamic CBCT reconstruction problem via different strategies. Cai et al. introduced low-rank matrix factorization into solving the dynamic CBCT, by viewing each temporal CBCT volume as a linear combination of a few basis images [34]. The linear coefficients and the basis images were solved simultaneously under a pre-defined matrix rank number (20). However, the study only reconstructed a single CBCT slice rather than the full 3D volume and only evaluated regular breathing scenarios. The low-rank assumption and the chosen rank number also remain to be further validated. Gao et al. viewed the 4D CBCT sequences as a product of spatial principal components and temporal motion coefficients [33]. Instead of solving the temporal motion coefficients directly from the angle-varying CBCT projections, their method proposes to learn the temporal motion coefficients from a previously-acquired 2D fluoroscopy sequence at a fixed gantry angle.

The learned *a priori* temporal motion coefficients from 2D projections were found to improve the CBCT reconstruction accuracy by reducing the degree of freedom in the spatiotemporal inverse problem. However, the described method relies on a motion trajectory learned from fixed-angle 2D projections, which may fail to represent the complex 3D motion and motion variations that occurred in the following CBCT acquisition. Taking 2D fluoroscopy images at fixed angles also incurs additional costs of imaging time and dose. The solved CBCT images are not fully time-resolved but are limited to 50 phases as well. Another study tried to solve dynamic CBCTs by combining projection-based motion estimation and motion-compensated reconstruction [35]. The method models the time kinematics via a series of time functions including surrogate motion signals. The motion irregularity and the poor representation/correlation of surrogate signals, however, may render the time regularization less effective and lead to motion estimation errors. The method was only evaluated on the parallel beam geometry as well.

Another category of methods introduces prior CT/CBCT images into solving the dynamic, time-resolved CBCTs [36]. They view each time-resolved CBCT as a deformed prior image via a deformation vector field (DVF). The principal component analysis (PCA) based method uses prior 4D-CT/4D-CBCT images to extract a patient-specific model of principal motion components and solves the DVF as a linear combination of the components [14, 36]. The substantial dimension reduction from PCA allows the linear coefficients to be solved from a single X-ray projection. However, a potentially major drawback of the pure DVF-driven CBCT estimation technique is that the variations between prior and new images may not be deformation alone [37]. The shading changes from different acquisition hardware (fan-beam CT vs cone-beam CT), various imaging protocols, and distinct noise/scatter patterns lead to errors when solving the motion fields. Non-deformation-induced anatomical changes and intensity variations can not be recovered by the DVFs [37]. Inter-scan deformations, such as tumor shrinkage, may not be captured by an intra-scan motion model like PCA either. A deep learning-based method was also developed to map cone-beam projections directly to PCA coefficients without explicit optimization [38]. It however suffers from similar issues as the conventional PCA-based techniques. Recently, another deep learning-based technique was proposed to directly convert single 2D projections into 3D volumes via a patient-specific encoder-decoder framework [39]. However, the 2D to 3D conversion technique is extremely ill-conditioned and its performance can be unstable to image intensity variations due to shading changes or noises. The two types of aforementioned deep learning methods also require a model to be pre-trained for each patient and each scan angle and may encounter generalizability issues when unseen motion occurs.

Recently, implicit neural representation (INR) learning has gathered many interests in the artificial intelligence field [40-42]. INR uses the power of neural networks, mostly multi-layer perceptrons (MLPs) [43], to construct and map complex objects including natural structures and medical images into continuous and differentiable functions. The MLPs can accept query inputs, for instance, the coordinates of image voxels, and output the physical properties like image intensities at queried voxels, to implicitly represent a complex medical image without specifying the details of the constitutive functions in advance. It offers a new way to reconstruct and represent volumetric objects and has recently been applied toward novel view synthesis, CT/MR reconstructions, and dose map compression [44-47]. A recent study also tried to use the INR to reconstruct a reference fan-beam CT volume, while using polynomially fitted motion fields solved via limited-angle projections to generate dynamic CT images from deforming the reference CT volume [48]. In this study, we proposed to use the representation capability of INR to develop a new dynamic CBCT reconstruction technique via simultaneous spatial and temporal INR learning (STINR). By STINR, we decoupled the complex spatiotemporal inverse problem of dynamic CBCT reconstruction into solving a spatial INR to represent a reference CBCT image, and several temporal INRs to represent the DVFs that characterize the time-resolved motion along different Cartesian directions. To reduce the complexity and leverage the inherent redundancy of DVFs, STINR combined PCA-based motion modeling with INR-based PCA coefficient learning to represent complex motion characteristics observed in each angle-varying projection. Compared to conventional machine/deep learning methods, STINR is a 'one-shot' learning technique that directly uses available cone-beam projection data to construct a patient-specific spatial and temporal imaging model that fits the specific projection set. In other words, the dynamic CBCT sequence is encoded by STINR as a neural network, which is solved on-the-fly in a self-supervised fashion with no prior training and no 'ground-truth' dynamic CBCTs required. Correspondingly, STINR does not suffer from the generalizability issue encountered by conventional pre-trained deep learning models [48]. In this study, we used the extended cardiac torso (XCAT) phantom to simulate dynamic volumetric images and projections of lung patients [49], featuring different regular and irregular motion scenarios, including motion amplitude/frequency variations, motion baseline shifts, and non-periodical motion. We also simulated different anatomical variation scenarios to represent inter-scan deformation, including tumor size shrinkage and tumor position change. We used STINR to reconstruct dynamic lung CBCTs, which were compared with the known 'ground-truth' lung CBCTs from the XCAT simulation and those generated by the conventional PCA-based method [14, 36]. We also compared STINR with the INR and polynomial fitting-based dynamic CT method as well [48]. The remaining paper is organized as follows: we first introduced the concept of INR and the simultaneous spatial and temporal learning of the STINR framework. Then the XCAT simulation details were introduced, including the various motion/anatomy scenarios. The results were presented and compared, followed by discussions of the advantages and limitations of the current methodology and evaluation.

## II. Materials and Methods:

STINR decoupled the dynamic CBCT reconstruction problem into the reconstruction of a reference CBCT volume ($CBCT_{Ref}$), and the simultaneous motion estimation by solving time-resolved DVFs $\boldsymbol{D}(t)$ to deform $CBCT_{Ref}$ to dynamic CBCTs ($CBCT_{dyn}$) at each time frame:

$$CBCT_{dyn} = CBCT_{Ref}\ (\boldsymbol{x} + \boldsymbol{D}(t)) \tag{1}$$

$\boldsymbol{x}$ denotes the voxel coordinates of $CBCT_{dyn}$, which were mapped to those of $CBCT_{Ref}$ through time-varying $\boldsymbol{D}(t)$ by trilinear interpolation [50]. Previous studies have found using a single MLP to map images both spatially and temporally to be particularly challenging [51], which yielded inaccurate results. STINR breaks a spatiotemporal INR into specialized partial INRs to represent the spatial and temporal kinematics independently, which reduces the overall complexity of the network and allows each component to be customized to fit the representation needs:

$$CBCT_{Ref} = \Phi^s(\theta) \tag{2}$$

$$\boldsymbol{D}(t) = \widetilde{\boldsymbol{PC}}_{dim,0} + \Phi^t_{dim,n}(\varphi) * \widetilde{\boldsymbol{PC}}_{dim,n}, \dim = v^1, v^2, v^3 \tag{3}$$

As shown in Eq. 2, the reference CBCT volume was represented by spatial INR $\Phi^s$ parameterized by the to-be-optimized coefficients $\theta$. In Eq. 3, $dim$ indicates the three Cartesian directions ($v^1, v^2, v^3$). Along each direction, a deformation matrix was defined. $n$ indicates the number of principal moton components along each dimension to model the deformation matrix. In this study, we used $n = 3$ as they were found sufficient to model the lung motion [52], while more can be readily applied. The temporal DVFs (deformation matrices) were constructed as scaled principal motion components ($\widetilde{PC}_{dim,n}$) by time-varying coefficients learned and represented as the temporal INRs $\Phi^t_{dim,n}$, which were parameterized by the to-be-optimized coefficients $\varphi$. $\widetilde{PC}_{dim,0}$ denotes the average DVF extracted from the PCA [14].

By decoupling the dynamic CBCT reconstruction problem into solving $CBCT_{Ref}$ and $D(t)$ separately, we reduced the complexity of the spatiotemporal inverse problem. The use of PCA-based motion modeling also introduces prior knowledge into solving the intra-scan motion, which helps to address the challenges of extreme under-sampling in dynamic CBCT reconstruction. By reconstructing $CBCT_{Ref}$ directly from on-board cone-beam projections, we avoided the challenges of shading mismatches and non-deformation-induced changes as encountered by methods that directly use prior images for registration [14, 36, 38]. The PCA motion model is only used to represent intra-scan motion while the inter-scan motion/deformation is implicitly solved via $CBCT_{Ref}$. The STINR thus enjoys unique advantages as compared to previously described methods. Below we first introduced the INR learning of $CBCT_{Ref}$, which is followed by the details of the INR learning of $D(t)$ that maps $CBCT_{Ref}$ into each temporal frame.

## II.A. Details of implicit neural representation learning of $CBCT_{Ref}$

### II.A.1. General methodology

As shown in Eq. 2 and Fig. 1, the INR-based reconstruction solves a function $\Phi^s$ to represent $CBCT_{Ref}$, in the form of a multi-layer perceptron. The MLP maps the 3-dimensional query voxel coordinates $x_i$ to their intensity distribution ($\sigma(x_i) \in \mathbb{R}^3$), where $i$ indicates the voxel number. The MLP-based function $\Phi^s$ is continuous, differentiable, and not limited to a specific voxel spatial resolution. It can be further defined as:

$$\Phi^s(x_i|\theta) = \tilde{\sigma}(x_i), x_i \in [-1,1]^3, \tilde{\sigma}(x_i) \in \mathbb{R}^3 \tag{4}$$

where $\tilde{\sigma}(x_i)$ denotes the output of the mapping function $\Phi^s$, which serves as the reconstruction and approximation of the true target property $\sigma(x_i)$. For the CBCT reconstruction problem, $\sigma(x_i)$ represents the attenuation coefficients of the scanned patient volume. The coordinates $x_i$ were normalized to the range [-1, 1] along each Cartesian direction, which was found to improve the INR learning accuracy [47].

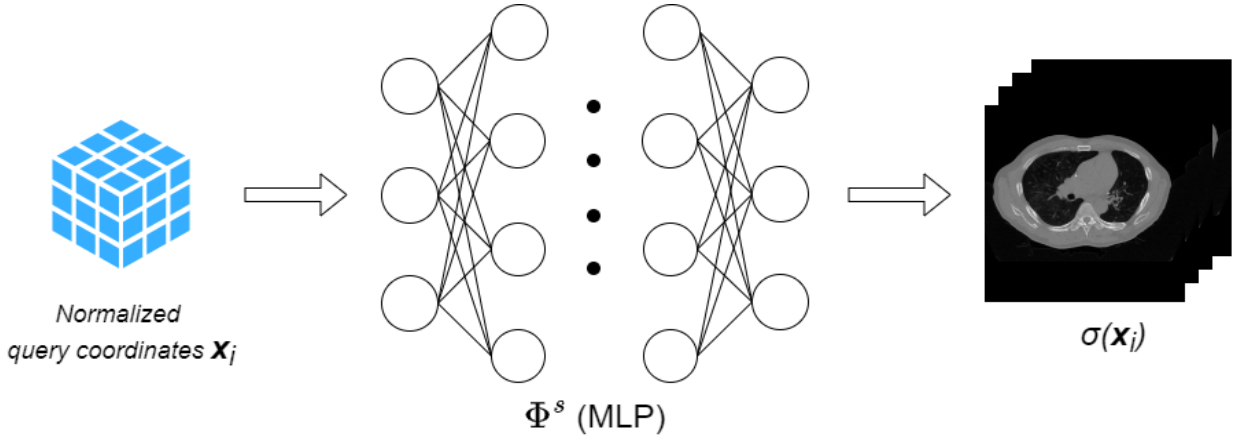

Figure 1. Scheme of the implicit neural representation of the reference CBCT volume, which uses a multi-layer perceptron (MLP, $\Phi^S$) to map query image coordinates ($x_i$) into the attenuation coefficients of the to-be-reconstructed CBCT ($\sigma$). The query coordinates were normalized to [-1, 1] along each Cartesian direction.

### II.A.2. Fourier feature encoding for the query coordinates

The vanilla coordinates-based MLPs were found difficult to learn high-frequency functions, as the neural tangent theory suggests the MLPs resemble kernels with rapid falloffs at high-frequency regions [53, 54]. To capture the high-frequency features in the images, the input query coordinates ($x_i$) of the neural network can be encoded by a large set of scalar functions before feeding into the MLP. Sinusoidal functions are commonly used to encode the query coordinates with Fourier features to fit the high-frequency signals. In this study, we used random Fourier feature (GRFF) encoding [55] which maps the input coordinates vector $x_i$ as:

$$\gamma(x_i) = [\sin(2\pi B x_i), \cos(2\pi B x_i)] \quad (5)$$

Where the matrix $B \sim \mathcal{N}(0, \sigma^2)$ is randomly sampled from a Gaussian distribution with width $\sigma$, which was determined empirically. In our implementation, we used 128 Fourier features for each input coordinate, with $\sigma = 2.5$. The encoded coordinates $\gamma(x_i)$ were fed subsequently into the MLP.

### II.A.3. Solving MLP for $CBCT_{Ref}$

The function of the MLP $\Phi^S$ is to map the coordinates $x_i$ to the true image intensity $\sigma(x_i)$ of $CBCT_{Ref}$, such that

$$\Phi^S(x_i|\theta) = \tilde{\sigma}(x_i) = \sigma(x_i) \quad (6)$$

Based on Eq. 6, the MLP parameters $\theta$ can be solved by minimizing a loss function defined as:

$$\theta = argmin_\theta L(\sum_{i=1}^{N} \Phi^S(x_i|\theta), \sigma(x_i)) \quad (7)$$

Where $L$ indicates the loss function between the reconstructed CBCT and the true CBCT. $N$ denotes all voxels within the CBCT volume. For reconstruction, the true attenuation coefficient map $\sigma(x_i)$ is not available to directly optimize the MLP, and only X-ray projections are provided. We can first reconstruct the X-ray projections into CBCT volumes using conventional analytical algorithms like the Feldkamp-Davis-Kress (FDK) algorithm [56], or other iterative algorithms [57, 58], and use the reconstructed volumes to replace $\sigma(x_i)$. Alternatively, the data fidelity can be optimized through a loss function directly defined on cone-beam projections like conventional iterative reconstruction algorithms:

$$\theta = argmin_\theta L(A \sum_{i=1}^{N} \Phi^S(x_i|\theta), P) \quad (8)$$

$P$ denotes the acquired cone-beam projections. $A$ denotes the system matrix which generates cone-beam projections from the CBCT volume, with the acquisition geometry identical to $P$. The loss function $L$ measures the distance between the forward projections of the INR-reconstructed CBCT volume and the true projections. In this study, we used the sum of squared differences as the distance metric:

$$\theta = argmin_\theta \left\| A \sum_{i=1}^{N} \Phi^s(x_i|\theta) - P \right\|_2^2 \qquad (9)$$

The parameters $\theta$ of the MLP can be conveniently optimized by minimizing the loss function. In our implementation, the MLP for $CBCT_{Ref}$ was constructed as four layers, with each layer containing 256 neurons. Except for the last layer, each layer was followed by a Swish activation function [59]:

$$actv(y) = y * sigmoid\,(y) = \frac{y}{1+e^{-y}} \qquad (10)$$

We compared Swish against Relu [60] and Siren [41] and found Swish provided slightly better results, which was chosen as the activation function in this study.

The above reconstructions and INR learning implicitly assume that the cone-beam projections $P$ contain only static projections without anatomical motion. For dynamic CBCT projections $P_t$, the underlying anatomy varies with time due to the physiological motion. Directly using these projections to reconstruct the spatial INR $\Phi^s$ will lead to motion artifacts-compromised images. To address this issue, the intra-scan DVFs $D(t)$, represented via the temporal INRs $\Phi^t$, are needed.

## II.B. Details of implicit neural representation learning of $D(t)$

Incorporating the intra-scan motion turns Eq. 9 into:

$$\theta, \varphi = argmin_{\theta,\varphi} \sum_t \| A\Phi^s(x + D(t) \mid \theta, \varphi) - P_t \|_2^2$$

$$= argmin_{\theta,\varphi} \sum_t \| A\Phi^s\big(x + \widetilde{PC}_{dim,0} + \Phi^t_{dim,n} * \widetilde{PC}_{dim,n} \mid \theta, \varphi\big) - P_t \|_2^2 \qquad (11)$$

In Eq. 11, we removed the subscript $i$ from $x_i$ to simplify the notation. With the principal components available, we can reconstruct $D(t)$ by estimating the weightings from the on-board projections and representing the weightings via the temporal INRs $\Phi^t_{dim,n}$. To obtain the principal components, we followed the previous works by performing inter-phase deformable registration within a planning 4D-CT volume [14, 36]. In radiotherapy, 4D-CTs are routinely acquired for sites impacted by respiratory motion and are widely available to provide high-quality prior knowledge [2]. We used the end-expiration (EE) phase volume as the reference volume and deformed it to the other phases to extract the inter-phase DVFs. The EE phase was selected due to its relative stability (and limited intra-phase motion) as compared to the other phases. The registration was performed using the open-source Elastix toolbox, of which the accuracy has been validated in many previous publications [61]. From these inter-phase DVFs, the eigenvectors (principal components) were extracted as $\widetilde{PC}_{dim,n}$ along each Cartesian dimension $(v^1, v^2, v^3)$. As previously mentioned, $\widetilde{PC}_{dim,0}$ was also extracted as the average of the inter-phase DVFs, denoting the DC component of the motion. For $n$, the first three PCA eigenvectors corresponding to the largest three eigenvalues were used since they are the most de-correlated and proved sufficient to model the lung motion [62].

In contrast to the spatial INR $\Phi^s$ which was fitted by one MLP, the temporal INRs $\Phi^t_{dim,n}$ were composed of 9 sub-MLPs, each of which represented one weighting for the three principal motion components along the three Cartesian directions. The temporal input, $t$, was normalized to [0, 1] and encoded by the GRFF (Eq. 5), like the spatial coordinate input, to enhance the INRs' capability to map high-frequency motion variations before being fed into each of the 9 MLPs. Same as the encoding used for spatial coordinates, we used 128 Fourier features with $\sigma = 2.5$. Since the temporal MLP only takes $t$ as input and the representation complexity is lower than the spatial MLP, each temporal MLP used 3 layers, with the first layer composed of 256 neurons, and the subsequent layers of 100 neurons each.

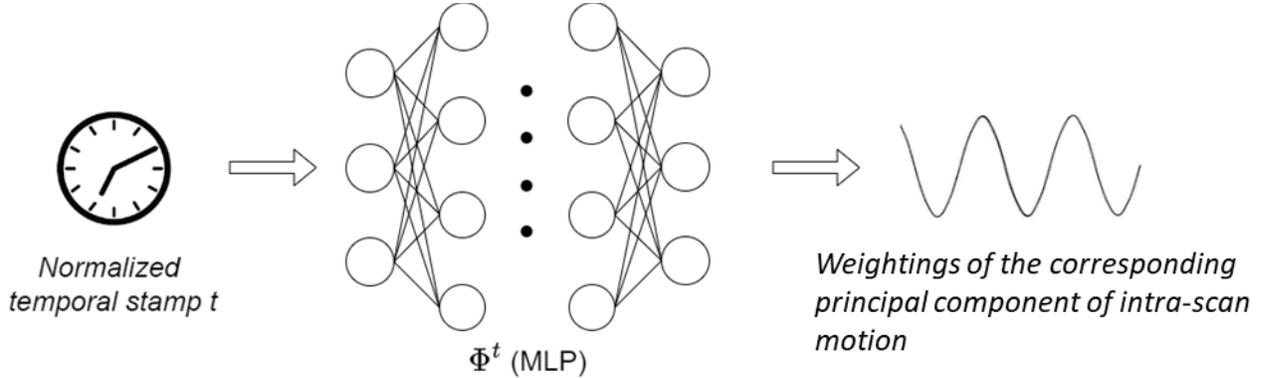

**Figure 2.** Scheme of the implicit neural representation of the intra-scan motion, which uses multi-layer perceptrons (MLP, $\Phi^t$) to map temporal time stamps ($t$) to the coefficients of the principal motion components. The temporal inputs were normalized to [0, 1]. In comparison to the single spatial INR, there were nine independent temporal INRs, each corresponding to one of the three principal components along each of the three Cartesian directions.

## II.C. The detailed workflow of STINR

Minimizing the objective function in Eq. 11 solves the spatiotemporal reconstruction problem of dynamic CBCTs. To further accelerate the INR learning process, and to reduce the possibility of the optimization being trapped at a local optimum for the ill-posed problem, we took a three-stage approach as shown in Fig. 3: (1). We initialized the reference volume INR $\Phi^s$ using a CBCT volume directly reconstructed by the Feldkamp-Davis-Kress (FDK) algorithm. Since the PCA motion model in II.B was derived based on the EE phase of the 4D-CT volume, we extracted the EE phase cone-beam projections from the full on-board projection set (Fig. 3) and reconstructed a coarse FDK volume. Using a reconstructed CBCT volume to directly solve the spatial INR $\Phi^s$ avoids the iterative forward-and-backward projection process and quickly pre-conditions the INR. (2). We further fine-tuned the reference volume INR $\Phi^s$ solved in (1) using the extracted cone-beam projections at the EE phase. To train the INR to better fit the true $CBCT_{Ref}$, this stage directly uses the EE projections to fine-tune the learned representation. Using projections directly can remove the artifacts introduced from the FDK reconstruction process. In both stage (1) and stage (2), the motion model was not introduced, assuming no intra-phase motion. In stage (3), the temporal INRs, along with the corresponding DVFs were introduced into mapping the $CBCT_{Ref}$ to dynamic $CBCT_{dyn}$. Digitally reconstructed radiographs (DRRs) were projected from $CBCT_{dyn}$, and compared with the acquired cone-beam projections $P_t$ to assess the representation learning loss (Eq. 11). At this stage, all the acquired cone-beam projections were used to compute and optimize the loss function. The spatial INR $\Phi^s$ and the temporal INRs $\Phi^t$ were optimized jointly and simultaneously during this stage. Through stage (3), the reference CBCT volume representation $\Phi^s$ was further corrected, to remove potential errors in stage (1) and stage (2) that assume no intra-phase motion. The time kinematics, as represented by $\Phi^t$, were fitted in free-form via the temporal INR learning towards all kinds of motion trajectories and scenarios.

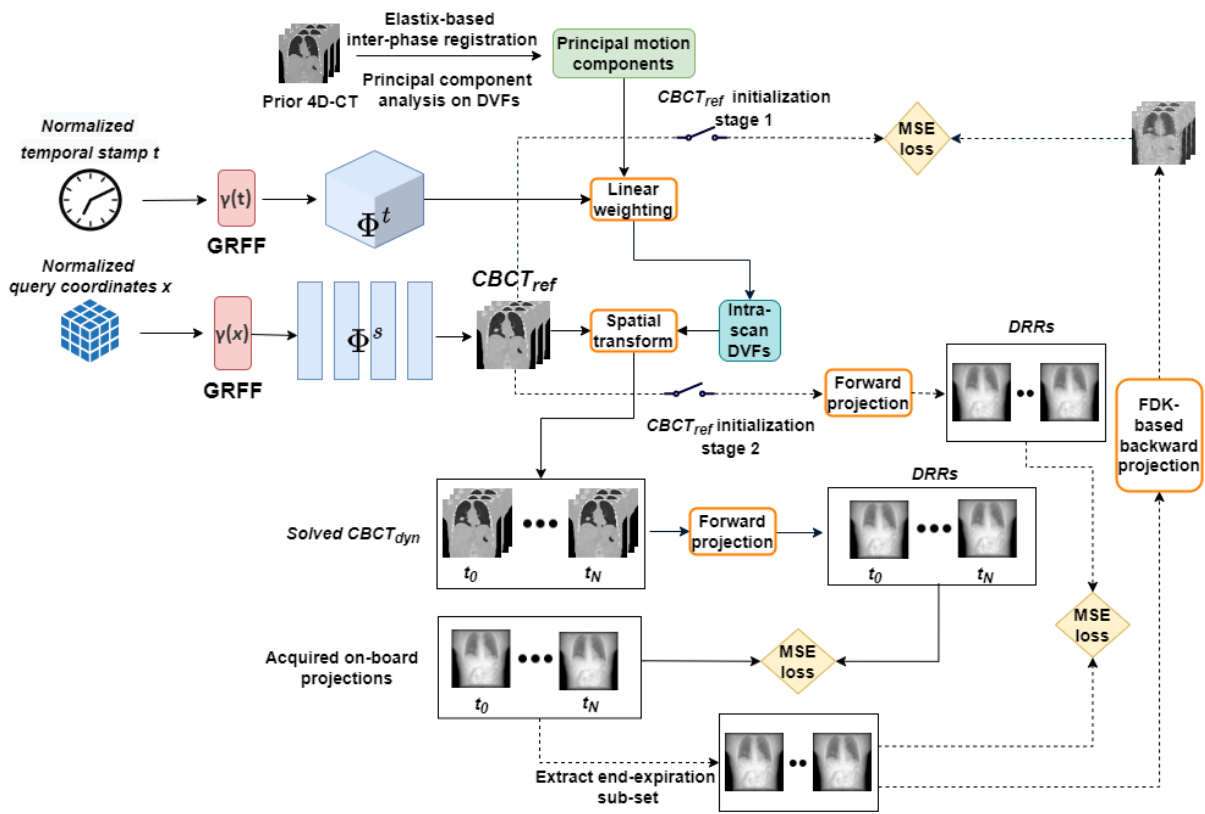

**Figure 3.** Detailed workflow of the overall STINR framework. The spatial and temporal inputs were encoded by the Fourier features and then fed into the corresponding MLPs to generate the reference CBCT volume ($CBCT_{Ref}$) and the principal component coefficients to derive dynamic intra-scan DVFs $\boldsymbol{D}(t)$. $\boldsymbol{D}(t)$ were applied to $CBCT_{Ref}$ to generate the dynamic CBCTs $CBCT_{dyn}$. The accuracy of $CBCT_{Ref}$ and $CBCT_{dyn}$ was evaluated either against reconstructed coarse CBCT volumes or by comparing digitally reconstructed radiographs (DRRs) of the reconstructed volumes with the acquired dynamic cone-beam projections.

In this study, we implemented the overall STINR framework based on the Pytorch backend (ver. 1.11.0). The optimization was performed automatically through the Pytorch framework. The learning rate was set to 0.002 for all the INRs. We used 500 iteration steps for stage 1, 500 iteration steps for stage 2, and 4000 iteration steps for stage 3.

### II.D. Experimental design

**II.D.1. Data curation**

To quantitatively evaluate the accuracy of the reconstructed dynamic CBCT volumes by STINR, in this study we used the 4D extended cardiac-torso (XCAT) phantom to simulate different breathing patterns and variations [49]. With XCAT, the reconstructed CBCT volumes can be directly compared with the 'ground-truth' simulated images for evaluation. The tracked target motion by the dynamic CBCTs can also be directly compared against the simulated motion trajectories. We simulated a total of eight motion/anatomy scenarios, featuring different patterns and degrees of motion variations/irregularities and anatomical changes (Table 1, Fig. 4).

**Table 1.** Details of the eight simulated motion/anatomy scenarios.

| Scenarios | Motion/Anatomy features |
|---|---|
| S1 | Small motion baseline shift |
| S2 | Large motion baseline shift |
| S3 | Motion frequency variations |
| S4 | Motion amplitude and baseline variations |
| S5 | Simultaneous motion frequency/amplitude variations |
| S6 | Non-periodic motion (or fast gantry rotation) |
| S7 | same as S1 but with the tumor diameter reduced by 50% from prior |
| S8 | same as S1 but with the tumor shifted from the prior position by 6 mm in each Cartesian direction |

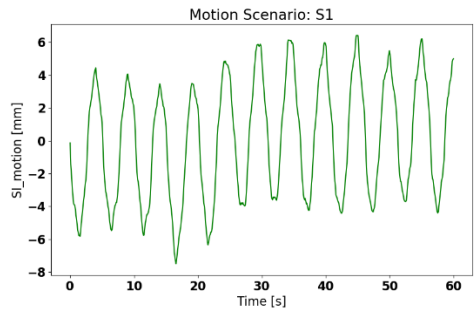
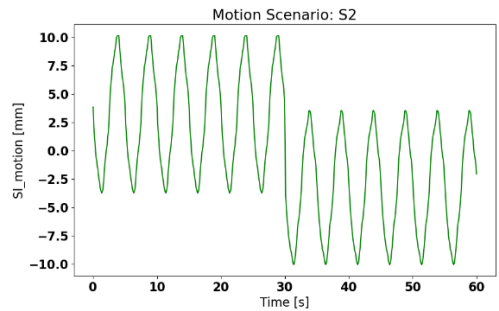

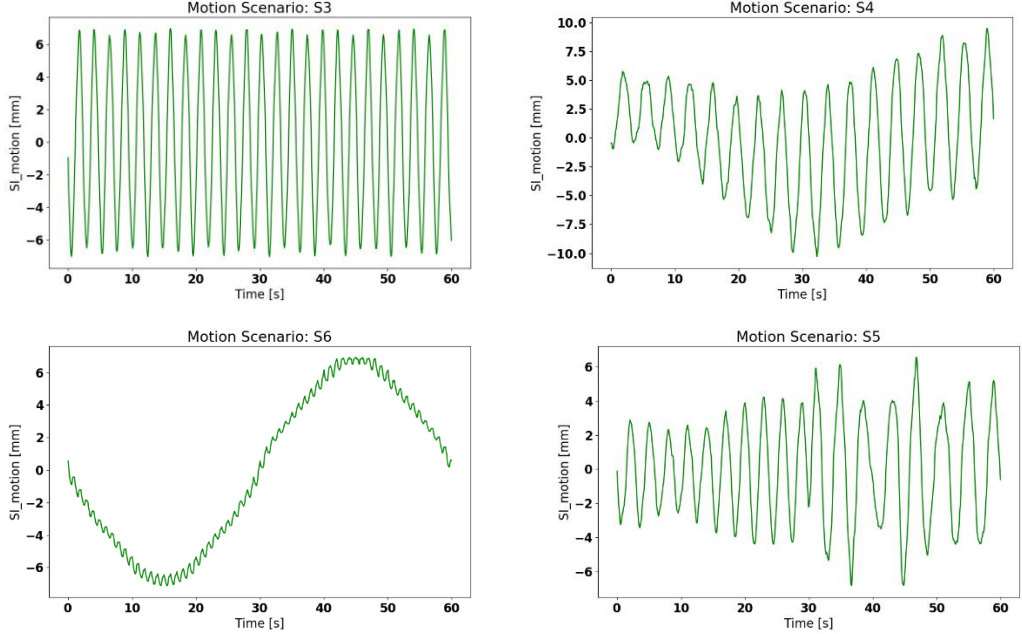

**Figure 4.** Simulated 'ground-truth' tumor motion trajectories corresponding to motion/anatomy scenarios S1-S6. S7 and S8 are not shown here due to their similarity to S1.

The baseline motion curve used by the XCAT simulation was a sinusoidal curve with a 5 s cycle, which was used to generate the 10-phase prior 4D-CT for the PCA motion modeling. For the simulated lung patient, we inserted a 15-mm spherical tumor in radius into the lower lobe of the right lung. As shown in Table 1 and Fig. 4, different on-board motion/anatomy variation scenarios during the CBCT acquisition were simulated, including motion baseline shift, motion frequency/amplitude variations, non-periodic motion, and tumor size and positional changes (simulating inter-scan anatomical variations). The non-periodic motion scenario (S6) is also similar to scenarios with extremely fast gantry rotation, such that the motion captured is not repeated (non-periodic). The 'ground-truth', dynamic on-board CBCT volumes were simulated of 1.5 mm x 1.5 mm x 1.5 mm spatial resolution per voxel, and of 256 x 256 x 256 voxels in dimension. For each motion/anatomy scenario, a dynamic CBCT volume was simulated every $\frac{1}{11}$ s, to match with the common frame rate (11 frames/s) used in clinical cone-beam projection acquisition scenarios [63]. We used a gantry rotation speed of 6°/s, which translates into a scan time of 60 s for a full 360° scan angle. In total, 660 ground-truth dynamic CBCT volumes were simulated for each motion/anatomy scenario. From each simulated dynamic volume, a corresponding cone-beam projection was simulated via the ray-tracing technique, using a gantry angle $\alpha$ defined in Eq. 12 based on the assumed gantry rotation speed and x-ray frame rate:

$$\alpha = \frac{1}{11} * 6 * (N-1), N = 1, 2 \ldots, 660 \qquad (12)$$

Here $N$ denotes the projection frame number under simulation. The CBCT projection was simulated with 512 x 512 pixels, with each pixel measuring 1.17 x 1.17 mm in dimension. The source-to-detector distance was 1500 mm and the source-to-isocenter distance was 1000 mm, and the gantry rotation axis was defined along the superior-inferior direction.

To implement STINR (Fig. 3), the cone-beam forward-projection layer and FDK back-projection layer were realized by using the differentiable ASTRA projectors [64] provided by the Operator Discretization Library (ODL)[65]. All computations were performed on an NVIDIA GeForce RTX 2080 super graphic processing unit (GPU) card with 8 GB memory. Due to the memory limit of the GPU card, we down-sampled the input cone-beam projections to 128 x 128 pixels in dimension, and the intermediate reconstruction volumes to 64 x 64 x 64 in dimension during the optimization.

**II.D.2. Comparison methods**

To benchmark the STINR technique against other currently available methods, we also evaluated the reconstruction results of the conventional PCA-based single-projection driven CBCT estimation technique ($PCA_{cv}$) [14, 36]. The objective of the conventional PCA method was formulated as:

$$w_{dim,n}^t = argmin_{w_{dim,n}^t} \|A\mu(x + D(t)) - P_t\|_2^2$$

$$= argmin_{w_{dim,n}^t} \|A\mu(x + \widetilde{PC}_{dim,0} + w_{dim,n}^t * \widetilde{PC}_{dim,n}) - P_t\|_2^2 \qquad (13)$$

Similar to STINR (Eq. 11), $PCA_{cv}$ uses the same principal motion components ($\widetilde{PC}_{dim,0}$, $\widetilde{PC}_{dim,n}$) for DVF derivation. However, instead of using a spatial INR to represent the reference CBCT volume, such a volume ($\mu$) of $PCA_{cv}$ was extracted from either a prior 4D-CT or could be reconstructed on-line from on-board projections. In this study, we used the EE-phase cone-beam projection subset (Fig. 3), extracted from all projections, to reconstruct a CBCT volume to serve as $\mu$. Since the PCA model was derived based on the EE phase of the prior 4D-CT (II.B), using the EE phase of cone-beam projections can maximize their similarity to fit the motion model, while also allowing inter-scan anatomical changes to be reconstructed and avoiding the potential shading variations among different imaging systems. To improve the reconstruction quality of the reference volume $\mu$, we used an algebraic reconstruction technique (ART) with the total variation regularization [66]. The ART-based image update (Eq. 14) and TV (Eq. 15) minimization steps were alternated until convergence is achieved:

$$\mu_{i+1} = \mu_i + \lambda a_{ij} \left[ \frac{p_j - \sum_i a_{ij}\mu_i}{\sum_i a_{ij}^2} \right] \qquad (14)$$

$$TV(\mu) = \|\nabla(\mu)\| \qquad (15)$$

Different from STINR which solves temporal INRs to model the temporal kinematics, $PCA_{cv}$ uses a scalar $w_{dim,n}^t$ to fit each principal component weighting at each temporal stamp (each projection), such that the optimization was performed independently per projection (Eq. 13). The objective function of Eq. 13 was analytically optimized using the non-linear conjugate gradient algorithm, of which the details could be found in our previous publications [14].

In addition to $PCA_{cv}$, we also compared STINR with the INR and polynomial fitting-based dynamic CT study ($INR_{poly}$) [48]. The $INR_{poly}$ method was developed to reconstruct dynamic CT images from limited-angle projections, with each temporal dynamic occupying a partial scan angle. The temporal DVFs were derived as voxel-wise motion coefficients weighted by temporal polynomials. To fit our reconstruction needs, we modified the $INR_{poly}$ method by introducing the cone-beam projection geometry and benchmarked its reconstruction results against STINR.

### II.D.3. Evaluation metrics

To quantitatively assess STINR and compare it against $PCA_{cv}$ and $INR_{poly}$, we evaluated the reconstructed dynamic CBCTs ($CBCT_{dyn}$) of each method by comparing with the ground-truth XCAT simulations via the relative error (RE) metric (Eq. 16). We also evaluated the solved intra-scan DVFs $\boldsymbol{D}(t)$, by comparing $\boldsymbol{D}(t)$-propagated lung tumor motion with the ground-truth tumor motion. We used the DICE coefficients (Eq. 16) and the center-of-mass errors (COMEs) [67] of the tumor contours.

$$RE = \sqrt{\frac{\sum(\mu_{recon} - \mu_{GT})^2}{\sum \mu_{GT}^2}} \quad (16)$$

$$DICE = 2 * \frac{|V_{recon} \cap V_{GT}|}{|V_{recon} + V_{GT}|} \quad (17)$$

In Eq. 16, $\mu_{recon}$ denotes the reconstructed dynamic CBCTs by different methods. $\mu_{GT}$ denotes the corresponding ground-truth images. The voxel-wise attenuation coefficient differences were computed and summed up to assess the overall reconstruction errors relative to the ground-truth attenuation coefficients. Eq. 17 defines the DICE coefficient, which measures the match between the dynamically resolved tumor volumes ($V_{recon}$) and the ground-truth tumor volumes ($V_{GT}$). For implementation, the tumors were manually segmented from the reconstructed reference CBCT volumes ($CBCT_{ref}$) for each combination of motion/anatomy scenario and reconstruction technique. The manual segmentations were propagated onto each dynamic volume using the DVFs $\boldsymbol{D}(t)$ solved by each technique, and then compared with the ground-truth tumor contours segmented via automatic intensity thresholding from the ground-truth dynamic CBCT volumes. A DICE coefficient of 1 indicates a perfect match and 0 indicates non-overlapping volumes. The COME metric, on the other hand, measures the distance between the dynamically resolved tumor location and the ground-truth tumor location, which also serves as an important metric to evaluate the accuracy of image guidance in radiotherapy.

## III. Results:

### III.A. Reference CBCT evaluation

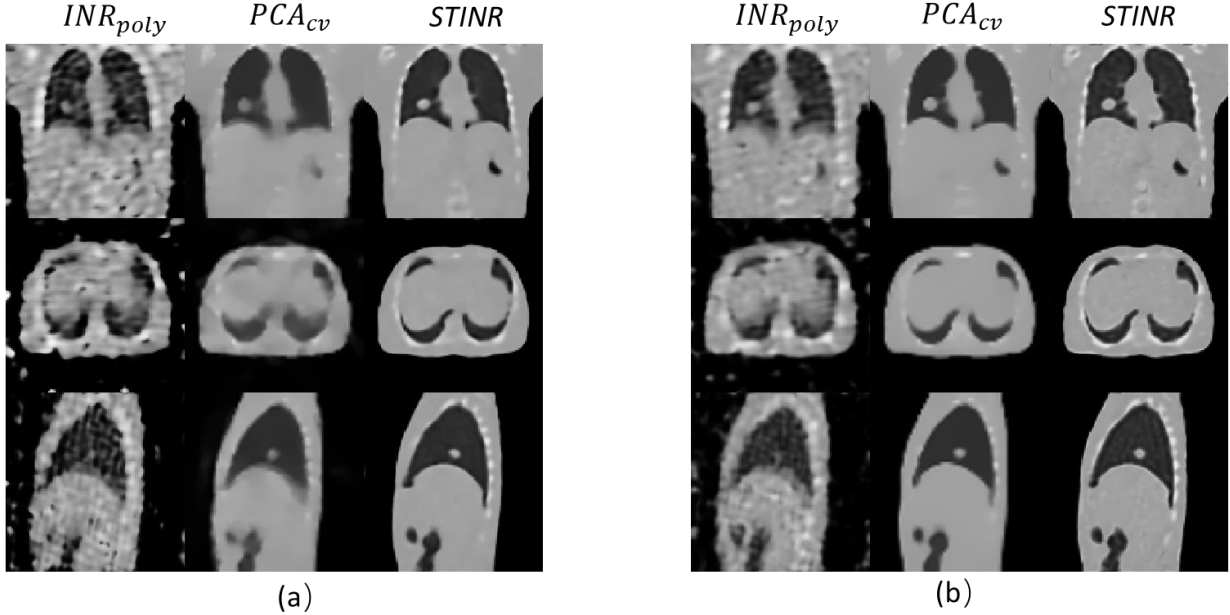

**Figure 5.** Comparison of the reconstructed reference CBCT volumes between $INR_{poly}$, $PCA_{cv}$, and STINR. The images shown here correspond to the motion/anatomy scenarios S2 (a) and S3 (b), respectively.

Fig. 5 compares the STINR-reconstructed reference CBCT volume $CBCT_{Ref}$ against those reconstructed by the $PCA_{cv}$ method and the $INR_{poly}$ method. For $PCA_{cv}$, the reference volume was reconstructed using the ART-based technique with the total variation regularization (Eq. 14 and Eq. 15). The images of Fig. 5 (a) correspond to the motion scenario S2 (Table 1), where a large intra-scan baseline shift was introduced. Due to the intra-scan baseline shift, $CBCT_{Ref}$ of $PCA_{cv}$ shows pronounced motion blurriness, since the EE phase projections extracted for $CBCT_{Ref}$ reconstruction contain two different motion baselines. In comparison, $CBCT_{Ref}$ of STINR achieves substantially more accurate reconstruction with the motion blurriness successfully subdued. The STINR images show sharp edges and well-defined tumor boundaries. Although $CBCT_{Ref}$ of STINR was also initialized using the extracted EE phase projections via the 3-stage optimization scheme (Fig. 3), the simultaneous spatial and temporal INR learning at stage 3 helps to incorporate the solved dynamic motion of each projection to correct the residual motion contained in $CBCT_{Ref}$. It also addresses potential motion mismatches between prior and new scans. With a more accurate $CBCT_{Ref}$, the dynamic DVFs can also be updated to represent more accurate motion for each projection. Compared to $PCA_{cv}$ and STINR, $INR_{poly}$ generally reconstructs the worst-quality reference image with amplified noise and motion blurriness. Without introducing prior knowledge from the PCA motion model, $INR_{poly}$'s accuracy was inferior due to the complexity of the spatiotemporal inverse problem, as the solution can be easily trapped at a local optimum.

Compared to Fig. 5 (a), Fig. 5 (b) corresponds to a scenario that only motion frequency variations were included (S3, Table 1). Without motion amplitude variations, correspondingly $CBCT_{Ref}$ of $PCA_{cv}$ was reconstructed with higher quality than that shown in Fig. 5 (a). However, its image quality is still inferior to that of STINR, as evidenced by the remaining motion blurriness and less well-defined bony structures. One contributing factor is that $PCA_{cv}$ can only use the sorted EE-phase projections to reconstruct $CBCT_{Ref}$, which only contains limited scan views. In contrast, $CBCT_{Ref}$ of STINR used all projections spanning the full scan views, enabling more accurate geometric and intensity reconstruction. The residual intra-phase motion from phase sorting contributed to $PCA_{cv}$'s inaccuracy as well.

## III.B. Dynamic CBCT evaluation

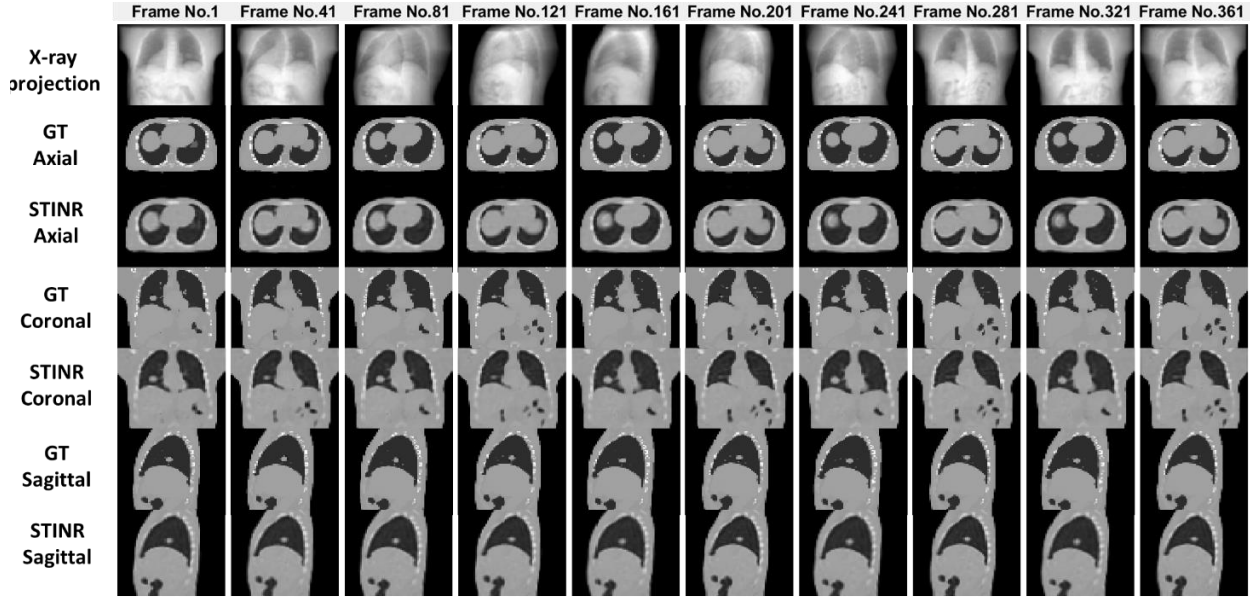

**Figure 6.** Comparison between the 'ground-truth' (GT) dynamic CBCTs and the STINR-reconstructed dynamic CBCTs at sampled x-ray projection frames, via three orthogonal views.

Fig. 6 shows the x-ray projections acquired at different time spots (frames, 1st row), the correspondingly reconstructed STINR dynamic CBCTs (axial view-3rd row, coronal view-5th row, and sagittal view-7th row), and the ground-truth dynamic CBCTs (axial view-2nd row, coronal view-4th row, and sagittal view-6th row). It can be observed that STINR can reconstruct dynamic CBCTs to match well with the ground-truth, in terms of both the image quality (boundary sharpness, intensity variations, noise level, etc.) and the fidelity of reconstructed anatomy (tumor location/shapes, lung, heart, air pockets, etc.).

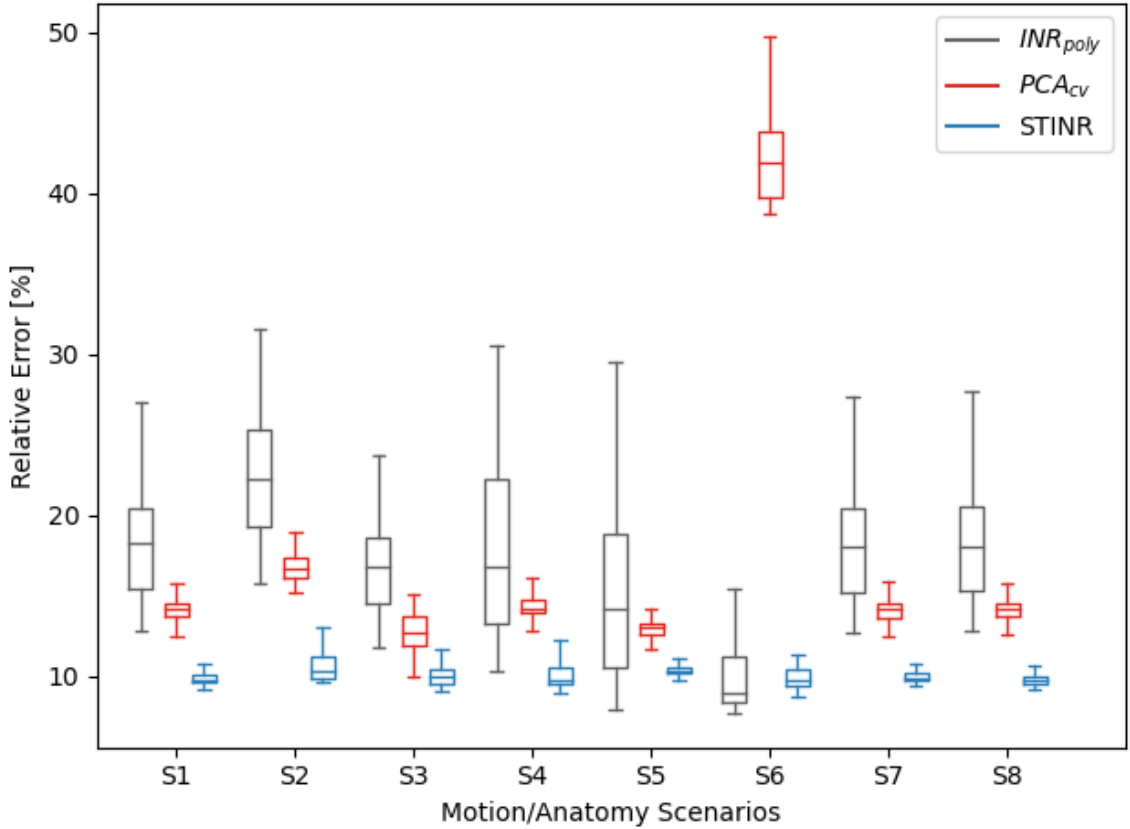

**Figure 7.** Boxplots of the relative error results for three different methods and all eight motion/anatomy scenarios.

**Table 2.** Mean and standard deviation of the relative errors (REs) for all motion/anatomy scenarios and three different methods. The RE was calculated between the reconstructed dynamic CBCT volume and the 'ground-truth' CBCT volume. Each motion scenario comprises 660 dynamic CBCT volumes.

| Motion scenarios | $INR_{poly}$ | $PCA_{cv}$ | STINR |
|---|---|---|---|
| S1 | 18.25 ± 3.26% | 14.06 ± 0.72% | 9.85 ± 0.47% |
| S2 | 22.39 ± 3.64% | 16.75 ± 0.83% | 10.57 ± 0.84% |
| S3 | 16.54 ± 2.63% | 12.58 ± 1.23% | 9.97 ± 0.53% |
| S4 | 17.75 ± 5.18% | 14.34 ± 0.77% | 10.08 ± 0.92% |
| S5 | 15.01 ± 5.19% | 12.88 ± 0.65% | 10.38 ± 0.42% |
| S6 | 10.40 ± 3.19% | 42.06 ± 2.53% | 9.85 ± 0.65% |
| S7 | 18.12 ± 3.39% | 14.05 ± 0.72% | 9.99 ± 0.43% |
| S8 | 18.25 ± 3.46% | 14.08 ± 0.72% | 9.79 ± 0.46% |

Fig. 7 presents the calculated REs between the reconstructed dynamic CBCTs by different methods. Each sub-boxplot corresponds to one motion/anatomy scenario (Table 1) and one reconstruction method. Similarly, STINR offers consistent reconstruction accuracy as compared to the other two methods. The $INR_{poly}$ performed poorly in general except for scenario S6. In contrast, $PCA_{cv}$ performed the worst for scenario S6, where non-periodic motion was simulated. Since $PCA_{cv}$ relies on the extracted EE-phase projections to reconstruct the reference CBCT volume, non-periodic motion (or equivalently fast gantry rotation) clustered these projections to a limited partial scan angle. It led to substantial anatomical and geometric distortions in the reconstructed reference CBCT volume, which were carried into the subsequently solved dynamic DVFs and CBCTs, and resulted in much worse dynamic reconstruction results for $PCA_{cv}$. In contrast, for $INR_{poly}$, the one-cycle motion of scenario S6 is easier to be fitted via polynomials as compared to the other scenarios, which resulted in relatively better performance. For STINR, since it could use all projections for simultaneous image and motion optimization, it was not susceptible to the limited-angle reconstruction errors of $PCA_{cv}$.

### III.C. Dynamic motion reconstruction evaluation

In addition to the images, we also evaluated the solved tumor motion by different methods. The comparison of tracked tumor motion curves along the superior-inferior (SI) direction was presented in Fig. 8, as the motion along SI direction was dominant. The corresponding DICE coefficient and COME results were reported in Table 3. As shown in Fig. 8 and Table 3, the STINR-solved tumor motion matched closely with the motion that was tracked from the ground-truth dynamic CBCT images. In general, the average DICE was above 0.8 for all scenarios except for scenario S7 (0.77) where the tumor diameter was reduced by 50%. The lower DICE values for S7 were expected due to the increased sensitivity of DICE to decreasing volumes. Similarly, the average COME was no larger than 2 mm for all scenarios, except for scenario S2 (2.4 mm) where a large baseline shift occurred. In comparison, $INR_{poly}$ failed almost completely to recover the high-frequency motion signals due to the limitation of the polynomial fitting approach. $PCA_{cv}$, on the other hand, also failed to correctly capture the true motion, especially for CBCTs close to motion peaks/valleys. It can be caused by the inferior quality of $CBCT_{Ref}$ for $PCA_{cv}$ (Fig. 5), as the poorly defined tumor/organ boundaries and shapes led to uncertainties in solving the PCA motion scaling factors. Echoing Fig. 7, $PCA_{cv}$ failed in solving scenario S6, due to the poor quality of $CBCT_{Ref}$ from limited-angle sampling.

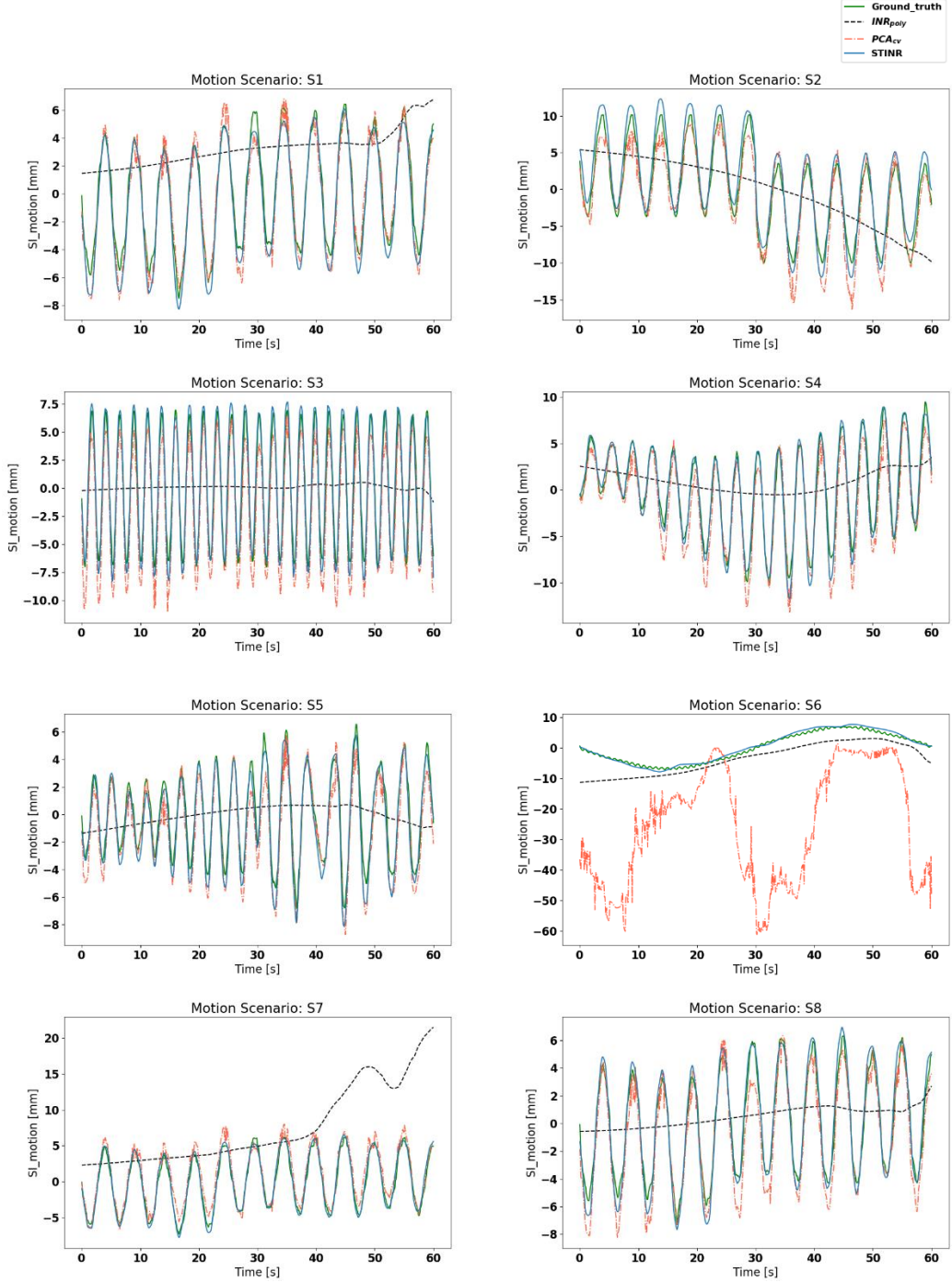

**Figure 8**. Comparison of tracked tumor motion along the superior-inferior (SI) direction for all reconstruction methods and motion/anatomy scenarios. The curves were style-coded and color-coded for differentiation (green solid: ground-truth; blue solid: STINR; red dot-dashed: $PCA_{cv}$; and black dashed: $INR_{poly}$).

**Table 3.** The DICE and COME results of dynamically resolved tumor volumes for all motion scenarios, by different methods. The DICEs and COMEs were calculated between the DVFs-propagated tumors and the 'ground-truth' tumor contours. Each motion scenario comprises 660 dynamic CBCT volumes.

| Motion scenarios | Evaluation metrics | $INR_{poly}$ | $PCA_{cv}$ | STINR |
|---|---|---|---|---|
| S1 | DICE | 0.73 ± 0.10 | 0.86 ± 0.03 | 0.87 ± 0.04 |
|    | COME [mm] | 5.6 ± 3.0 | 1.5 ± 0.7 | 1.4 ± 0.6 |
| S2 | DICE | 0.69 ± 0.10 | 0.65 ± 0.06 | 0.85 ± 0.04 |
|    | COME [mm] | 6.9 ± 3.4 | 2.9 ± 1.8 | 2.4 ± 1.0 |
| S3 | DICE | 0.68 ± 0.10 | 0.84 ± 0.05 | 0.86 ± 0.03 |
|    | COME [mm] | 6.4 ± 2.9 | 2.1 ± 1.0 | 2.0 ± 0.7 |
| S4 | DICE | 0.72 ± 0.09 | 0.83 ± 0.06 | 0.88 ± 0.04 |
|    | COME [mm] | 5.6 ± 3.1 | 3.2 ± 1.8 | 1.6 ± 0.8 |
| S5 | DICE | 0.79 ± 0.09 | 0.78 ± 0.40 | 0.84 ± 0.04 |
|    | COME [mm] | 4.0 ± 2.4 | 3.1 ± 0.8 | 1.2 ± 0.6 |
| S6 | DICE | 0.78 ± 0.07 | 0.24 ± 0.25 | 0.84 ± 0.04 |
|    | COME [mm] | 3.3 ± 2.3 | 30.8 ± 20.1 | 1.6 ± 0.6 |
| S7 | DICE | 0.47 ± 0.13 | 0.75 ± 0.06 | 0.77 ± 0.06 |
|    | COME [mm] | 9.2 ± 5.3 | 2.8 ± 0.7 | 1.6 ± 0.8 |
| S8 | DICE | 0.77 ± 0.07 | 0.84 ± 0.04 | 0.87 ± 0.04 |
|    | COME [mm] | 4.5 ± 2.3 | 1.6 ± 0.8 | 1.3 ± 0.5 |

## IV. Discussion:

Dynamic CBCT imaging generates volumetric images with superior temporal and spatial resolutions and is highly desired in clinical applications including radiotherapy targeting verification, tumor motion monitoring and prediction, treatment dose tracking and accumulation, and robust treatment planning. Our study developed a simultaneous spatial and temporal implicit neural representation learning (STINR) framework to address the extremely challenging problem of reconstructing dynamic CBCTs from singular X-ray projections. In comparison to traditional voxel-wise representations of volumetric images, the STINR method uses different multi-layer perceptrons to represent the spatial features and temporal evolutions of imaged anatomical structures. MLP is powerful in representing images or motions with complex, continuous, and differentiable functions, while needless to specify/constrain the detailed function forms in advance. Compared to conventional voxel-wise image representations, MLP maps the image/motion to neural networks and in theory, can generate voxelized moving images of arbitrary spatial or temporal resolutions to achieve inherent super-resolution.

Due to the difficulty of mapping the full spatiotemporal imaging series into a single MLP, we developed STINR to use individual MLPs to fit the spatial and temporal INRs separately, which allows the MLPs to be customized to tailor to the inherent complexity variations of different representation tasks (imaging/motion). We introduced patient-specific prior knowledge, the PCA-based motion models, into solving the temporal INRs to further reduce the complexity of the ill-conditioned spatiotemporal inverse problem. PCA-based motion modeling allows substantial dimension reductions to represent complex motion scenarios accurately and effectively. Conventional PCA-based dynamic CBCT reconstruction methods solve the motion based on a fixed reference CBCT volume, which can be reconstructed from on-board projections [68], or extracted from patient-specific prior 4D-CTs [14]. The reference CBCT volume is subsequently used to solve the PCA motion coefficients to construct dynamic motion. In comparison, STINR reconstructs the reference CBCT volume as a spatial INR and optimizes the reference CBCT volume (spatial INR) and the dynamic DVFs (temporal INRs) simultaneously. The simultaneous solution of spatial and temporal INRs allows STINR to correct the residual intra-phase motions of the reference CBCT volume, which can come from sources including projection sorting errors, motion baseline shifts, or other irregular motion (Fig. 4). For non-periodic motion or fast gantry rotation scenarios, it can be rather challenging to reconstruct such a reference volume for conventional PCA-based methods, since the projections of a certain motion phase may only occupy a restricted small angle instead of scattering around the full scan angle as in periodic motion or slow gantry rotation scenarios. The limited scan angle severely distorts the reconstructed reference volume and significantly reduces the accuracy of dynamic CBCT reconstruction for conventional PCA-based methods (S6 in Table 2, Table 3, Fig. 7, Fig. 8). In comparison, STINR uses all dynamic projections to resolve the reference volume and the intra-scan motion simultaneously, which allows the reference volume to be reconstructed based on information from all projection angles while accounting for the intra-scan motion. The full-angle coverage successfully removes the limited-angle distortions to generate a high-quality reference volume, which in return helps to solve intra-scan DVFs more accurately and reliably.

In this study, we used the on-board projections at the EE phase to reconstruct reference CBCT volumes for the conventional PCA-based reconstruction technique, since the PCA motion model is as well built on the EE phase volume of the prior 4D-CT. Alternatively, previous studies also directly used the EE phase volume of the prior 4D-CT as the reference volume [14, 36]. The challenge of this approach is that it cannot fully resolve the inter-scan anatomical variations due to factors like treatment response (tumor shrinkage, lung inflammation, etc.) or disease progression, which cannot be learned from an intra-scan motion model of PCA as these long-term variations do not exist or happen within a single 4D-CT scan. Another challenge of this approach is from the shading variations between CT/CBCT, as the prior 4D-CT is usually acquired in a fan-beam geometry while the latter CBCT is acquired in cone-beam geometry. The cone-beam geometry usually suffers from amplified degradation signals including photon scatter and electronic noise [66, 69]. The difference in x-ray source, energy, mA, ms, and other hardware variations also introduce difficulties/inaccuracy in directly using the CT volume as the reference for on-board dynamic CBCT reconstruction.

In addition to the conventional PCA-based dynamic CBCT reconstruction methods, in this study, we also compared STINR with another INR method using polynomial fitting to generate the intra-scan DVFs [48]. In comparison to STINR, the polynomial fitting-based INR method was originally developed for limited-angle reconstruction problems, where the motion is expected to be slow and gradual (such that each motion state will occupy at least a small scan angle). Our dynamic CBCT reconstruction problem, however, is much more challenging due to the motion scenarios evaluated (much faster and more volatile breathing motion). Results show that the polynomial fitting-based method fails to accurately capture the breathing motion that occurred within most dynamic CBCT scans. Compared to the temporal INRs, the polynomials showed very limited accuracy in representing periodical motion and its variations. Another advantage of STINR over the polynomial fitting-based method is the introduction of PCA-based motion modeling as prior knowledge. The polynomial fitting-based INR method needs to solve PCA-like motion matrices from the dynamic projections directly, which incurred substantially elevated complexity and trapped the reconstructions into obvious sub-optimal solutions in our evaluations.

In our study, we used XCAT to simulate dynamic patient volumes and cone-beam projections, which allowed us to control and customize the motion/anatomical scenarios to test the accuracy and robustness of STINR with known 'ground-truth'. Previous studies also frequently used XCAT to generate dynamic volumes to test their algorithms [33, 34]. However, although XCAT provides highly anthropomorphic features, it is still warranted to use real patient cone-beam projections and images to validate STINR. The challenge of using real patient data, however, is attaining ground-truth dynamic volumetric images and motion to validate our methods, which may be an impossible task. One workaround is to compare the projected image features from the reconstructed dynamic CBCTs to those extracted from the dynamic cone-beam projections, for instance, the locations of radiopaque fiducial markers or directly tracked anatomical features [38]. In addition, studies using anthropomorphic physical phantoms are also warranted.

In this study, we have to convert the continuously-defined INRs into intermediate voxelized images to feed into forward/backward ASTRA operators to drive optimization. Although the INRs allow CBCTs to be represented at arbitrary resolutions due to their continuous nature, due to the limit of the GPU memory, we have to down-sample the volume matrix dimensions to 64 x 64 x 64 voxels during the optimization, which yielded a 6 mm spatial resolution alone each Cartesian direction. Although the final output CBCTs can be up-scaled to arbitrarily high spatial resolution, the intermediate down-sampling may adversely affect the final reconstruction accuracy. In our study, we found a residual DVF error of around 2 mm (Table 6), which could be partially caused by the above-mentioned spatial resolution limitations. Further investigations quantifying the effects of such under-sampling operations, and the improvements from using larger memory GPUs or low memory footprint operators, are warranted to examine the potential and limitations of STINR.

The current STINR framework solves the reference CBCTs and dynamic DVFs without implementing additional spatial and temporal regularizations. Introducing further *a priori* knowledge, for instance, the total variation regularization on the spatial domain [37], may further improve the reconstruction accuracy. Regularizations in the motion domain, like those encouraging the smoothness of motion, may also be applied in caution (not to substantially suppress the reconstruction of irregular motion). Another way is to further condition the STINR framework and initialize the temporal INRs via tracked motion curves, for instance, those from the external surface marker tracking or internal diaphragm tracking [35]. The pre-conditioning may help to further accelerate the reconstruction speed and reduce the chances of the reconstructions being trapped at local optima. In addition to the temporal INR initialization, the spatial INR can also be initialized by prior CT/CBCT images, which was found effective in improving image reconstruction accuracy while reducing the on-board imaging sampling needs [46]. However, potential biases from the spatial and temporal INR initializations, especially due to large anatomical variations (spatial) or the imperfect correlations between the surrogate motion and the real volumetric motion (temporal), may negatively affect the reconstruction accuracy and remain to be further evaluated.

## V. Conclusion:

We developed a spatial and temporal implicit neural representation learning method (STINR) to reconstruct dynamic CBCTs from singular X-ray projections acquired during a normal 60 s CBCT scan. The 'one-shot' learning STINR method uses the powerful representation capability of multi-layer perceptrons and adopts prior knowledge from PCA-based motion modeling. The combination successfully addresses the challenging spatiotemporal inverse problem of dynamic CBCT reconstruction. The STINR can be easily adapted and implemented for other imaging modalities including magnetic resonance imaging. The spatially- and temporally- resolved images can benefit many clinical applications, including target motion tracking/prediction, intra-treatment dose calculation/accumulation, and robust planning/delivery. Future studies based on real patient data with various motion/anatomy scenarios are warranted for a comprehensive evaluation and potential methodology fine-tuning.

## VI.  Acknowledgments

The study was supported by funding from the National Institutes of Health (R01CA240808, R01CA258987). We would like to thank Dr. Hua-Chieh Shao at the UT Southwestern Medical Center for helpful discussions regarding the methodology implementation. We also would like to thank Dr. Paul Segars at Duke University for providing the XCAT phantom for our study.

# References

1. Jaffray, D.A., et al., *Flat-panel cone-beam computed tomography for image-guided radiation therapy.* International Journal of Radiation Oncology* Biology* Physics, 2002. **53**(5): p. 1337-1349.
2. Pan, T., et al., *4D-CT imaging of a volume influenced by respiratory motion on multi-slice CT.* Medical physics, 2004. **31**(2): p. 333-340.
3. Létourneau, D., et al., *Cone-beam-CT guided radiation therapy: technical implementation.* Radiotherapy and oncology, 2005. **75**(3): p. 279-286.
4. Pereira, G.C., M. Traughber, and R.F. Muzic, *The role of imaging in radiation therapy planning: past, present, and future.* BioMed research international, 2014. **2014**.
5. Borst, G.R., et al., *Kilo-voltage cone-beam computed tomography setup measurements for lung cancer patients; first clinical results and comparison with electronic portal-imaging device.* International Journal of Radiation Oncology* Biology* Physics, 2007. **68**(2): p. 555-561.
6. Topolnjak, R., et al., *Breast patient setup error assessment: comparison of electronic portal image devices and cone-beam computed tomography matching results.* International Journal of Radiation Oncology* Biology* Physics, 2010. **78**(4): p. 1235-1243.
7. Kong, V.C., A. Marshall, and H.B. Chan, *Cone beam computed tomography: the challenges and strategies in its application for dose accumulation.* Journal of Medical Imaging and Radiation Sciences, 2016. **47**(1): p. 92-97.
8. Sibolt, P., et al., *Clinical implementation of artificial intelligence-driven cone-beam computed tomography-guided online adaptive radiotherapy in the pelvic region.* Physics and imaging in radiation oncology, 2021. **17**: p. 1-7.
9. Zhang, Y., F.F. Yin, and L. Ren, *Dosimetric verification of lung cancer treatment using the CBCTs estimated from limited-angle on-board projections.* Medical physics, 2015. **42**(8): p. 4783-4795.

10. Liang, J., et al., *Intrafraction 4D-cone beam CT acquired during volumetric arc radiotherapy delivery: kV parameter optimization and 4D motion accuracy for lung stereotactic body radiotherapy (SBRT) patients.* Journal of Applied Clinical Medical Physics, 2019. **20**(12): p. 10-24.
11. Song, W.Y., et al., *A dose comparison study between XVI® and OBI® CBCT systems.* Medical physics, 2008. **35**(2): p. 480-486.
12. Sonke, J.J., et al., *Respiratory correlated cone beam CT.* Medical physics, 2005. **32**(4): p. 1176-1186.
13. Vergalasova, I., J. Maurer, and F.F. Yin, *Potential underestimation of the internal target volume (ITV) from free-breathing CBCT.* Medical physics, 2011. **38**(8): p. 4689-4699.
14. Zhang, Y., et al., *A technique for estimating 4D-CBCT using prior knowledge and limited-angle projections.* Medical physics, 2013. **40**(12): p. 121701.
15. Zhang, Y., et al., *Preliminary clinical evaluation of a 4D-CBCT estimation technique using prior information and limited-angle projections.* Radiotherapy and Oncology, 2015. **115**(1): p. 22-29.
16. Thengumpallil, S., et al., *Difference in performance between 3D and 4D CBCT for lung imaging: a dose and image quality analysis.* Journal of applied clinical medical physics, 2016. **17**(6): p. 97-106.
17. Sweeney, R.A., et al., *Accuracy and inter-observer variability of 3D versus 4D cone-beam CT based image-guidance in SBRT for lung tumors.* Radiation Oncology, 2012. **7**(1): p. 1-8.
18. Bergner, F., et al., *Autoadaptive phase-correlated (AAPC) reconstruction for 4D CBCT.* Medical physics, 2009. **36**(12): p. 5695-5706.
19. Li, T. and L. Xing, *Optimizing 4D cone-beam CT acquisition protocol for external beam radiotherapy.* International Journal of Radiation Oncology* Biology* Physics, 2007. **67**(4): p. 1211-1219.
20. Shieh, C.C., et al., *SPARE: Sparse-view reconstruction challenge for 4D cone-beam CT from a 1-min scan.* Medical physics, 2019. **46**(9): p. 3799-3811.
21. Yan, H., et al., *A hybrid reconstruction algorithm for fast and accurate 4D cone-beam CT imaging.* Medical physics, 2014. **41**(7): p. 071903.
22. Harris, W., et al., *Estimating 4D-CBCT from prior information and extremely limited angle projections using structural PCA and weighted free-form deformation for lung radiotherapy.* Medical physics, 2017. **44**(3): p. 1089-1104.
23. Wang, J. and X. Gu, *Simultaneous motion estimation and image reconstruction (SMEIR) for 4D cone-beam CT.* Medical physics, 2013. **40**(10): p. 101912.
24. Huang, X., et al., *U-net-based deformation vector field estimation for motion-compensated 4D-CBCT reconstruction.* Medical physics, 2020. **47**(7): p. 3000-3012.
25. Leng, S., et al., *High temporal resolution and streak-free four-dimensional cone-beam computed tomography.* Physics in medicine & biology, 2008. **53**(20): p. 5653.
26. Clements, N., et al., *The effect of irregular breathing patterns on internal target volumes in four-dimensional CT and cone-beam CT images in the context of stereotactic lung radiotherapy.* Medical physics, 2013. **40**(2): p. 021904.
27. Pan, C.H., et al., *The irregular breathing effect on target volume and coverage for lung stereotactic body radiotherapy.* Journal of applied clinical medical physics, 2019. **20**(7): p. 109-120.
28. Huang, L., et al., *A study on the dosimetric accuracy of treatment planning for stereotactic body radiation therapy of lung cancer using average and maximum intensity projection images.* Radiotherapy and Oncology, 2010. **96**(1): p. 48-54.
29. Yasue, K., et al., *Quantitative analysis of the intra-beam respiratory motion with baseline drift for respiratory-gating lung stereotactic body radiation therapy.* Journal of Radiation Research, 2022. **63**(1): p. 137-147.

30. Cooper, B.J., et al., *Quantifying the image quality and dose reduction of respiratory triggered 4D cone-beam computed tomography with patient-measured breathing.* Physics in Medicine & Biology, 2015. **60**(24): p. 9493.
31. Poulsen, P.R., et al., *Kilovoltage intrafraction motion monitoring and target dose reconstruction for stereotactic volumetric modulated arc therapy of tumors in the liver.* Radiotherapy and Oncology, 2014. **111**(3): p. 424-430.
32. Li, X., et al., *Effects of irregular respiratory motion on the positioning accuracy of moving target with free breathing cone-beam computerized tomography.* International journal of medical physics, clinical engineering and radiation oncology, 2018. **7**(2): p. 173.
33. Gao, H., et al., *Principal component reconstruction (PCR) for cine CBCT with motion learning from 2D fluoroscopy.* Medical physics, 2018. **45**(1): p. 167-177.
34. Cai, J.-F., et al., *Cine cone beam CT reconstruction using low-rank matrix factorization: algorithm and a proof-of-principle study.* IEEE transactions on medical imaging, 2014. **33**(8): p. 1581-1591.
35. Jailin, C., et al., *Projection-based dynamic tomography.* Physics in Medicine & Biology, 2021. **66**(21): p. 215018.
36. Li, R., et al., *Real-time volumetric image reconstruction and 3D tumor localization based on a single x-ray projection image for lung cancer radiotherapy.* Med Phys, 2010. **37**(6): p. 2822-6.
37. Zhang, Y., et al., *A new CT reconstruction technique using adaptive deformation recovery and intensity correction (ADRIC).* Med Phys, 2017. **44**(6): p. 2223-2241.
38. Wei, R., et al., *Real-time tumor localization with single x-ray projection at arbitrary gantry angles using a convolutional neural network (CNN).* Physics in Medicine & Biology, 2020. **65**(6): p. 065012.
39. Shen, L., W. Zhao, and L. Xing, *Patient-specific reconstruction of volumetric computed tomography images from a single projection view via deep learning.* Nature biomedical engineering, 2019. **3**(11): p. 880-888.
40. Eslami, S.M.A., et al., *Neural scene representation and rendering.* Science, 2018. **360**(6394): p. 1204-+.
41. Sitzmann, V., et al., *Implicit neural representations with periodic activation functions.* Advances in Neural Information Processing Systems, 2020. **33**: p. 7462-7473.
42. Peng, S., et al. *Neural body: Implicit neural representations with structured latent codes for novel view synthesis of dynamic humans*. in *Proceedings of the IEEE/CVF Conference on Computer Vision and Pattern Recognition*. 2021.
43. Heidari, A.A., et al., *Ant lion optimizer: theory, literature review, and application in multi-layer perceptron neural networks.* Nature-inspired optimizers, 2020: p. 23-46.
44. Sitzmann, V., M. Zollhöfer, and G. Wetzstein *Scene Representation Networks: Continuous 3D-Structure-Aware Neural Scene Representations*. 2019. arXiv:1906.01618.
45. Lombardi, S., et al. *Neural Volumes: Learning Dynamic Renderable Volumes from Images*. 2019. arXiv:1906.07751.
46. Shen, L., J. Pauly, and L. Xing, *NeRP: implicit neural representation learning with prior embedding for sparsely sampled image reconstruction.* IEEE Transactions on Neural Networks and Learning Systems, 2022.
47. Vasudevan, V., et al., *Implicit neural representation for radiation therapy dose distribution.* Physics in Medicine & Biology, 2022. **67**(12): p. 125014.
48. Reed, A.W., et al. *Dynamic ct reconstruction from limited views with implicit neural representations and parametric motion fields*. in *Proceedings of the IEEE/CVF International Conference on Computer Vision*. 2021.
49. Segars, W.P., et al., *4D XCAT phantom for multimodality imaging research.* Med Phys, 2010. **37**(9): p. 4902-15.

50. Bourke, P., *Interpolation methods.* Miscellaneous: projection, modelling, rendering, 1999. **1**(10).
51. Four Dimensional, C. and R.Y. Hirano, *Four Dimensional CT Using Implicit Neural Representation.*
52. Li, R., et al., *On a PCA-based lung motion model.* Physics in Medicine & Biology, 2011. **56**(18): p. 6009.
53. Basri, R., et al., *Frequency Bias in Neural Networks for Input of Non-Uniform Density.* International Conference on Machine Learning, Vol 119, 2020. **119**.
54. Tancik, M., et al., *Fourier features let networks learn high frequency functions in low dimensional domains.* Advances in Neural Information Processing Systems, 2020. **33**: p. 7537-7547.
55. Tancik, M., et al. *Fourier Features Let Networks Learn High Frequency Functions in Low Dimensional Domains*. 2020. arXiv:2006.10739.
56. Feldkamp, L.A., L.C. Davis, and J.W. Kress, *Practical cone-beam algorithm.* Josa a, 1984. **1**(6): p. 612-619.
57. Andersen, A.H. and A.C. Kak, *Simultaneous algebraic reconstruction technique (SART): a superior implementation of the ART algorithm.* Ultrasonic imaging, 1984. **6**(1): p. 81-94.
58. Wang, J., T. Li, and L. Xing, *Iterative image reconstruction for CBCT using edge-preserving prior.* Medical physics, 2009. **36**(1): p. 252-260.
59. Ramachandran, P., B. Zoph, and Q.V. Le, *Searching for activation functions.* arXiv preprint arXiv:1710.05941, 2017.
60. Agarap, A.F., *Deep learning using rectified linear units (relu).* arXiv preprint arXiv:1803.08375, 2018.
61. Klein, S., et al., *elastix: a toolbox for intensity-based medical image registration.* IEEE Trans Med Imaging, 2010. **29**(1): p. 196-205.
62. Ruan, D. and P. Keall, *Online prediction of respiratory motion: multidimensional processing with low-dimensional feature learning.* Phys Med Biol, 2010. **55**(11): p. 3011-25.
63. Ling, C., et al., *Acquisition of MV-scatter-free kilovoltage CBCT images during RapidArc™ or VMAT.* Radiotherapy and Oncology, 2011. **100**(1): p. 145-149.
64. van Aarle, W., et al., *Fast and flexible X-ray tomography using the ASTRA toolbox.* Optics Express, 2016. **24**(22): p. 25129-25147.
65. J. Adler, H.K., and O. Oktem. *Operator discretization library (ODL)*. 2017; Available from: https://github.com/odlgroup/odl.
66. Ouyang, L., K. Song, and J. Wang, *A moving blocker system for cone-beam computed tomography scatter correction.* Medical physics, 2013. **40**(7): p. 071903.
67. Zhang, Y., et al., *4D liver tumor localization using cone-beam projections and a biomechanical model.* Radiotherapy and Oncology, 2019. **133**: p. 183-192.
68. Dhou, S., et al., *Fluoroscopic 3D Image Generation from Patient-Specific PCA Motion Models Derived from 4D-CBCT Patient Datasets: A Feasibility Study.* Journal of Imaging, 2022. **8**(2): p. 17.
69. Chen, X., et al., *Optimization of the geometry and speed of a moving blocker system for cone-beam computed tomography scatter correction.* Medical physics, 2017. **44**(9): p. e215-e229.